\documentclass{aastex}
\usepackage{emulateapj5}

\usepackage{amssymb}
\usepackage{amsmath}
\usepackage{graphicx}

\newcommand{\bfv}{{\boldsymbol{v}}}

\slugcomment{Received 2001 October 11; accepted 2001 November 6}
\shorttitle{Dark matter halos around ellipticals}
\shortauthors{Baes \& Dejonghe}

\begin{document}

\title{Dark matter halos around elliptical galaxies: how reliable\\ is the
stellar kinematical evidence?}

\author{Maarten Baes\altaffilmark{1} and Herwig Dejonghe}

\affil{Sterrenkundig Observatorium, Universiteit Gent, Krijgslaan
281-S9, B-9000 Gent, Belgium;}

\email{maarten.baes@rug.ac.be, herwig.dejonghe@rug.ac.be}

\altaffiltext{1}{Postdoctoral Fellow of the Fund for Scientific
Research, Flanders, Belgium (FWO-Vlaanderen)}

\begin{abstract}
Hierarchical models of galaxy formation and various observational
evidence suggest that elliptical galaxies are, like disk galaxies,
embedded in massive dark matter halos. Stellar kinematics are
considered the most important tracer for this dark halo at a few
effective radii. Using detailed modeling techniques, several
authors have recently presented stellar kinematical evidence of a
dark halo for a number of elliptical galaxies. In these modeling
techniques, dust attenuation (absorption and scattering of
starlight by dust grains) has not been taken into account.
Nevertheless, elliptical galaxies contain a significant amount of
interstellar dust, which affects all observable quantities,
including the observed kinematics. We constructed a set of
dynamical models for elliptical galaxies, in which dust
attenuation is included through a Monte Carlo technique. We find
that a dust component, shallower than the stellar distribution and
with an optical depth of order unity, affects the observed
kinematics significantly, in the way that it mimics the presence
of a dark halo. If such dust distributions are realistic in
elliptical galaxies, we are faced with a new mass-dust degeneracy.
Taking dust attenuation into account in dynamical modeling
procedures will hence reduce or may even eliminate the need for a
dark matter halo at a few effective radii.
\end{abstract}


\section{Introduction}

It has been generally accepted for decades that disk galaxies are
ubiquitously embedded in massive dark matter halos. During the
past few years, a consensus has been developing that elliptical
galaxies also contain dark halos. Their existence is predicted by
hierarchical theories of galaxy formation, and has recently been
supported by various observational evidence, such as gravitational
lensing (Griffiths et al.\ 1996; Keeton, Kochanek, \& Falco 1998)
and X-ray measurements of their hot gas atmospheres (Matsushita et
al.\ 1998; Loewenstein \& White 1999). Although useful to infer
the large-scale mass distribution of elliptical galaxies, these
observations do not sufficiently constrain the detailed structure
of the mass distribution at a few effective radii. This region is
particularly important for understanding the coupling of the dark
and luminous matter. In order to constrain the gravitational
potential at these radii, other, kinematical, tracers are used.
For disk galaxies, the neutral hydrogen gas, which radiates at 21
cm, forms an excellent kinematical tracer; elliptical galaxies,
however, usually lack the necessary amounts of interstellar gas.
Discrete tracers such as planetary nebulae, globular clusters and
dwarf satellite galaxies can be used (Zepf et al.\ 2000;
Romanowski \& Kochanek 2001; Kronawitter et al.\ 2000), but due to
their small numbers and larger distances to the center, they do
not sufficiently constrain the gravitational potential at a few
$R_e$. This leaves stars as the main tracer for the mass
distribution in elliptical galaxies in this region.

The first stellar kinematical evidence for dark matter halos
around elliptical galaxies came in the early 1990s. At that time,
the available kinematical data consisted of the mean projected
velocity $\langle v_p \rangle$ and the projected velocity
dispersion $\sigma_p$, at projected radii rarely larger than $1
R_e$. For a number of elliptical galaxies, the projected velocity
dispersion profile was found to decrease only slowly with
projected radius, a behavior that was interpreted as a signature
for the presence of a dark matter halo (Saglia, Bertin, \&
Stiavelli 1992; Saglia et al.\ 1993). However, such $\sigma_p$
profiles can also be generated by intrinsically tangentially
anisotropic galaxy models, without the need for dark matter halos.
The velocity dispersion profile alone does not contain sufficient
kinematic information to constrain both the mass and the orbital
structure of elliptical galaxies, a problem usually referred to as
the mass-anisotropy degeneracy (Gerhard 1993). This degeneracy can
be broken by considering the higher order kinematical information
contained in the line-of-sight velocity distributions (LOSVDs),
usually parameterized by means of the Gauss-Hermite shape
parameters $h_i$ (where $i\geq3$; Gerhard 1993; van der Marel \&
Franx 1993). In particular, the additional information contained
in the $h_4$ profile provides the key to breaking the
mass-anisotropy degeneracy: the combination of a slowly decreasing
velocity dispersion profile, together with a relatively large
$h_4$ profile, is generally interpreted as evidence for a dark
matter halo.\footnote{For a detailed analysis of the effects of
mass and anisotropy on the projected dispersion and $h_4$
profiles, we refer to section~3 of Gerhard et al.\ (1998).} Thanks
to improved instrumentation and data reduction techniques, it is
nowadays possible to determine $\langle v_p \rangle$, $\sigma_p$,
$h_3$ and $h_4$ profiles with a reasonable accuracy, out to
several $R_e$. Several authors have recently adopted such
kinematical information to constrain the dark matter distribution
in a number of elliptical galaxies (Rix et al.\ 1997; Gerhard et
al.\ 1998; Saglia et al.\ 2000; Kronawitter et al.\ 2000).

In this entire discussion, it has (implicitly) been assumed that
elliptical galaxies consist of two dynamically important
components: the stars moving as test particles in a gravitational
potential, generated by both stellar and dark mass. During the
past decade, it has become well established that elliptical
galaxies also contain a surprisingly large amount of interstellar
dust (up to several million solar masses), most of it believed to
be distributed diffusely over the galaxy (Roberts et al.\ 1991;
Goudfrooij \& de Jong 1995; Bregman et al.\ 1998). This number
must be revised an order of magnitude upward, if more detailed
dust mass estimators (Merluzzi 1998) or additional submillimeter
measurements (Fich \& Hodge 1993; Wiklind \& Henkel 1995) are
taken into account. This is still a negligible fraction of the
total mass of the galaxy, such that the dust will hardly influence
the gravitational potential. Nevertheless, it has a significant
role in galaxy dynamics: dust grains efficiently absorb and
scatter optical photons. Interstellar dust will therefore affect
all observable quantities, including the observed kinematics.

We are undertaking an effort to understand the impact of
interstellar dust on the observed kinematics in elliptical
galaxies. Previously, we investigated how absorption by dust
grains affects the light profile and the observed kinematics (Baes
\& Dejonghe 2000; Baes, Dejonghe, \& De Rijcke 2001). We found
that the observed kinematics are affected only in the most central
regions, the magnitude of these effects being of the order of a
few percents. Here we have extended our modeling to incorporate
the process of scattering off dust grains. We will show in this
Letter that this has a considerable effect on the observed
kinematics, in particular concerning the stellar kinematical
evidence for dark matter halos.

\section{The modeling}

To investigate the effects of attenuation\footnote{We will refer
to attenuation as the combined effect of absorption and
scattering} on the observed kinematics, we constructed a
spherically symmetric elliptical galaxy model, consisting of a
stellar and a dust component. For the stellar distribution, we
adopted an isotropic Hernquist model. It is a reasonable
approximation for the observed surface brightness distribution of
elliptical galaxies, and has the advantage that most of the
internal stellar kinematics can be calculated analytically
(Hernquist 1990). The distribution of the (diffuse) dust in
elliptical galaxies is not well constrained; its presence has been
demonstrated only indirectly. We adopted a dust distribution that
is shallower than the stellar density, with a total visual optical
depth of order unity. Such a model seems to be indicated by the
observed color gradients in elliptical galaxies (Witt et al.\
1992; Goudfrooij \& de Jong 1995; Wise \& Silva 1996). For the
optical properties of the dust grains (the relative opacity, the
scattering albedo, and the phase function), we adopted the values
calculated by Maccioni \& Perinotto (1994) and displayed in Di
Bartolomeo, Barbaro \& Perinotto (1995). A more extended set of
models, with a larger variety in optical depth and star-dust
geometry, will be presented in a forthcoming paper (M. Baes \& H.
Dejonghe, 2001, in preparation).

The inclusion of dust absorption in the calculation of the LOSVDs
was rather straightforward: it required an extra factor in the
integration along the line of sight (Baes \& Dejonghe 2000). If
scattering is included, however, such an approach is not possible
anymore. Indeed, photons can now leave their initial path, and
reach the observer through a completely different path. Therefore,
we adopted a Monte Carlo technique to include this process.
Basically, the method consists of following the individual path of
a very large number of photons (typically about $10^7$), selected
randomly from phase space, according to the phase space
distribution function of the galaxy. Knowledge of the path along
which each photon leaves the galaxy, together with the kinematical
signature of the star that emitted it, enables us to construct the
surface brightness profiles and LOSVDs of the galaxy, both with
and without dust attenuation taken into account. The $\sigma_p$
and $h_4$ profiles are extracted from the LOSVDs (the $\langle
v_p\rangle$ and $h_3$ profiles are identically zero, because we
consider a spherically symmetric nonrotating galaxy model). A more
detailed description of the method will be given in M. Baes \& H.
Dejonghe (2001, in preparation).

A check on the method was provided by the case where only
absorption is taken into account: we found the results of the
Monte Carlo simulation to be in perfect agreement with our
previously obtained semianalytical results (Baes \& Dejonghe
2000).

\section{The observed kinematics}

In Figure 1, the effect of dust attenuation on the observed
kinematics is illustrated. For the lines of sight that pass near
the galaxy center, the kinematics are nearly unaffected. Because
of the modest (and conservative) dust masses we used, the
high-velocity stars in the galaxy center are only slightly
obscured. As a result, the velocity dispersion only decreases
marginally. At large radii, however, the kinematics are seriously
affected. The $\sigma_p$ profile drops less steeply, and the $h_4$
parameter is significantly larger compared to the dust-free case.
In Figure~2, we show the LOSVD at $R=5 R_e$ of our galaxy model,
with and without dust attenuation. This figure clearly indicates
that dust attenuation brings on significant high-velocity wings in
the outer LOSVDs. In particular, it should be noted that these
LOSVDs do not vanish at the local escape velocity: hence, stars
are observed that can physically not be present on these lines of
sight. These wings are a scattering effect, as demonstrated in
Figure 3, and they are responsible for the increase of $\sigma_p$
and $h_4$. Attenuation by interstellar dust apparently has a
kinematical signature that is strikingly similar to the presence
of a dark matter halo: a velocity dispersion profile that
decreases more slowly than expected, and a relatively large $h_4$
profile. Hence, dust-affected kinematics mimic the presence of a
dark matter halo.

\section{Modeling the dust-affected kinematics}

To check this into more detail, we considered our dust-affected
kinematics as an observational data set, and modeled it as any
dynamical modeler would do, i.e., without taking dust attenuation
into account. We concentrated on the model with $\tau_V=1$, and
considered a data set consisting of the $K$-band photometry and
$V$-band $\sigma_p$ and $h_4$ profiles, out to $5 R_e$. The
modeling was performed with a powerful non-parametric modeling
technique based on quadratic programming (Dejonghe 1989).

\subsection{Models with a constant $M/L$}

First, we tried to find out whether the data were consistent with
a model with a constant $M/L$. We constructed a set of dynamical
models, with $M/L$ as a free parameter. The best-fitting model is
represented by the dashed line in Figure 4. Obviously, this fit is
not satisfactory: it can fit the $\sigma_p$ profile, but only
through a strong tangential anisotropy, reflected in a strongly
negative $h_4$ profile. It is impossible to fit both the
$\sigma_p$ and $h_4$ profiles with a constant $M/L$ model.

\subsection{Models with a dark matter halo}

In order to construct models with a rising $M/L$, i.e.\ with a
dark halo, we considered a set of Hernquist potentials, each with
a different scale length. We constructed a set of dynamical
models, with both the potential scale length and the mass-to-light
ratio as free parameters. The best fit to the photometry and the
$\sigma_p$ and $h_4$ profiles\footnote{The slight disagreement at
the center of the h4 profile is unimportant, as we are primarily
interested in the outer regions.} is plotted in solid lines in
Figure 4. Its potential has a scale length 45\% larger than the
original Hernquist potential, and in the maximum stellar mass
hypothesis (the analogue for the maximum disk hypothesis in disk
galaxies), the dark matter contributes roughly a third of the
total mass within 1\,$R_e$, and half of the total mass within the
last data point. This clearly demonstrates that the effects of
dust attenuation can mimic the presence of a dark matter halo.
Taking dust attenuation into account in dynamical modeling
procedures will hence reduce or may even eliminate the need for a
dark matter halo at a few $R_e$.

\section{Conclusions}

In view of these results, we may have to reconsider the stellar
kinematical evidence for dark matter halos. In analogy with the
mass-anisotropy degeneracy, which could be lifted by considering
higher-order shape parameters for the LOSVDs, we are now faced
with a new degeneracy, which could be called the mass-dust
degeneracy. Indeed, the attenuation (in particular the scattering)
by dust grains has the same effect on the stellar kinematics as a
dark matter halo. At least, these results apply for the model we
have explored in this Letter, with a dust distribution shallower
than the stars and with an optical depth $\tau_V=1$. We are well
aware that these assumptions are quite uncertain, and an extended
set of models, with a large variety in optical depth and star-dust
geometry, is being investigated.

The new mass-dust degeneracy strongly complicates the use of
stellar kinematics as a tracer for the mass distribution in
elliptical galaxies. Although the presence of dark matter halos
nowadays seems firmly established at very large scales, this
leaves us with a major problem concerning the determination of the
dark halo properties at a few effective radii. There are two
possible ways to break this new degeneracy. The first option is to
observe the kinematics at near-infrared wavelengths, where the
effects of dust are negligible. This, however, poses a serious
observational challenge in the off-center regions. The second
option is to include radiative transfer calculations in dynamical
modeling techniques. Besides a computational challenge, this
requires a better knowledge of the spatial distribution and of the
optical properties of dust grains in elliptical galaxies than we
have today. The new generation of far-infrared and submillimeter
instruments, such as ALMA and {\em SIRTF}, will help to solve this
problem.

\acknowledgements

The authors would like to thank T.\ van Albada and A.\ N.\ Witt
for stimulating discussions, and the referee, N. Cretton, for his
constructive remarks. M.\ B.\ acknowledges the financial support
of FWO-Vlaanderen.

\begin{figure}
\centering
\includegraphics[clip]{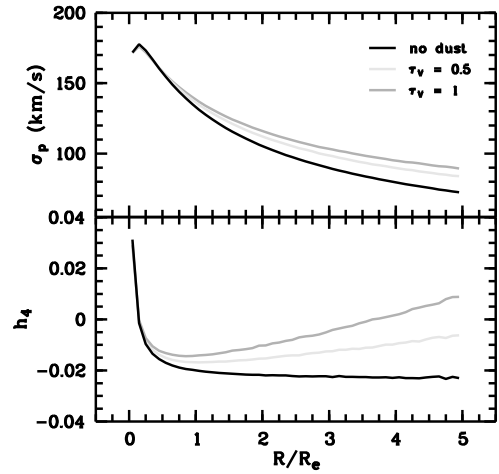}
\caption{Effect of dust attenuation on the projected kinematics,
in particular the $\sigma_p$ and the $h_4$ profiles. The profiles
are shown for a model without dust attenuation and for models with
optical depths of $\tau_V=0.5$ and $\tau_V=1$.}
\end{figure}

\begin{figure}
\centering
\includegraphics[clip]{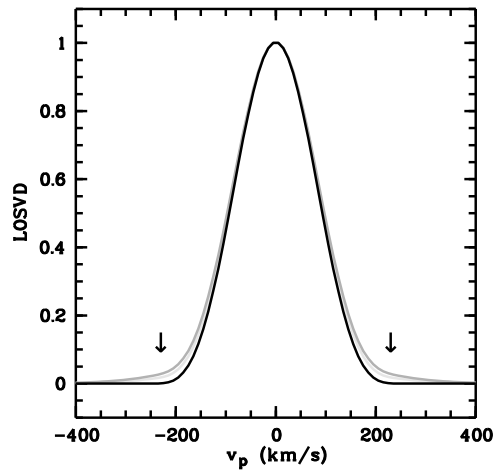}
\caption{Effect of dust attenuation on the outer LOSVDs. Shown is
the LOSVD at $R=5 R_e$; the different curves correspond to those
in Fig.\ 1. The LOSVDs are normalized to unity at $v_p=0$. The
escape velocity $v_{\text{esc}}$ at this line of sight is
indicated by an arrow. If dust attenuation is not taken into
account, no stars are observed with line-of-sight velocities
larger than $v_{\text{esc}}$. If dust attenuation is considered,
however, the LOSVD does not vanish at $v_{\text{esc}}$.}
\end{figure}

\begin{figure}
\centering
\includegraphics[clip,width=65mm]{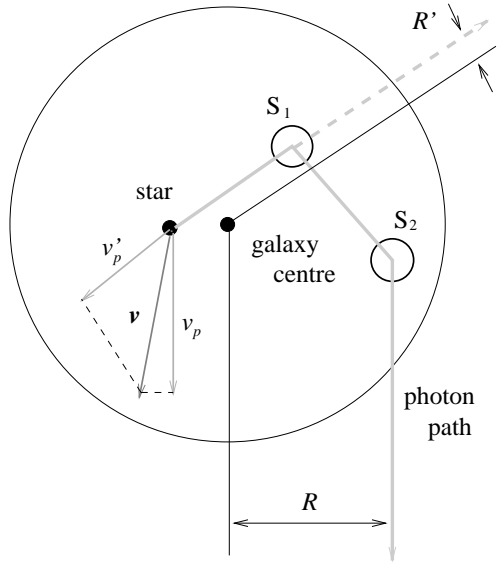}
\caption{Scattering as the reason for the high velocity wings in
the outer LOSVDs. A photon is emitted by a star near the galaxy
center, which has an intrinsic velocity $\bfv$. If there were no
dust attenuation, the photon would travel along a straight line,
and contribute a line-of-sight velocity $v_p'$ to the LOSVD at the
projected radius $R'$. In a dusty galaxy, however, photons can
leave their initial path and reach the observer through a
completely different path. In the figure, the photon is scattered
twice, at $S_1$ and $S_2$, and it leaves the galaxy along a line
of sight with projected radius $R$. During its journey through the
galaxy, the photon carries along the kinematical signature of the
star that emitted it. Hence, it will contribute the line-of-sight
velocity $v_p$ to the LOSVD at the line of sight $R$. In
particular, it is possible that the observed line-of-sight
velocity $v_p$ exceeds the local escape velocity
$v_{\text{esc}}(R)$. This is indeed observed in the outer LOSVDs:
they do not vanish at the local escape velocity (Fig.\ 2).}
\end{figure}

\begin{figure}
\centering
\includegraphics[clip]{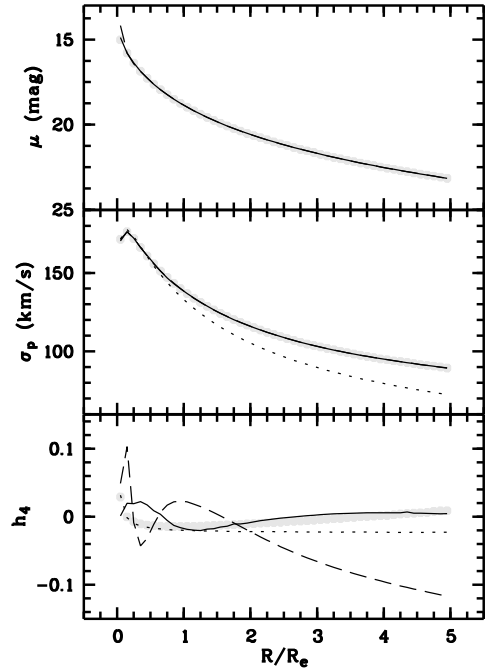}
\caption{Two fits to the photometric and kinematical data
corresponding to the model with optical depth $\tau_V=1$. Shown
are the surface brightness $\mu$, velocity dispersion $\sigma_p$,
and $h_4$ profiles. The data are represented by the gray circles.
The dashed line is the best fitting model with constant $M/L$; the
solid line is the best fitting model with a dark matter halo (see
text). The projected kinematics of the input model, without dust
attenuation taken into account, are represented by a dotted line.}
\end{figure}

\end{document}